\documentclass[twocolumn,nofootinbib]{aastex6}
\usepackage{graphicx}	
\usepackage{amsmath}
\usepackage{amssymb}
\usepackage{subfigure}
\usepackage{color}
\usepackage{txfonts}
\usepackage{tabto}
\usepackage{braket}
\usepackage{lineno}
\newcommand\mceq{{\sc MCEq}}

\newcommand\simlt{\lower.5ex\hbox{$\; \buildrel < \over \sim \;$}}
\newcommand\simgt{\lower.5ex\hbox{$\; \buildrel > \over \sim \;$}}

\begin{document}
\title{Polarized radiation and the Emergence of Biological Homochirality on Earth and Beyond
}
\author{No\'emie Globus\altaffilmark{1,2} Anatoli Fedynitch\altaffilmark{3} and Roger D. Blandford\altaffilmark{4}}
\altaffiltext{1}{Center for Cosmology \& Particle Physics, New-York University, New-York, NY10003, USA. E-mail:  globus@nyu.edu}
\altaffiltext{2}{Center for Computational Astrophysics, Flatiron Institute, Simons Foundation, New-York, NY10003, USA}
\altaffiltext{3}{Institute for Cosmic Ray Research, the University of Tokyo, 5-1-5 Kashiwa-no-ha, Kashiwa, Chiba 277-8582, Japan. E-mail: afedyni@icrr.u-tokyo.ac.jp}
\altaffiltext{4}{Kavli Institute for Particle Astrophysics \& Cosmology, Stanford University, Stanford, CA 94305, USA. E-mail:  rdb3@stanford.edu}

\begin{abstract}
It has been proposed that spin-polarized cosmic radiation  can  induce asymmetric changes in helical biopolymers that may account for  the emergence of biological homochirality. 
 The parity violation in the weak interaction has direct consequences on the transport of polarization in cosmic ray showers.
 In this paper, we show that muons retain their polarization down to energies at which they can initiate enantioselective mutagenesis. Therefore, muons are  most likely to succeed in establishing the connection between broken symmetries in the standard model of particle physics  and that found in living organisms. 
 We calculate the radiation doses deposited by primary and secondary cosmic rays at various prime targets for the searches of life in the solar system: Mars, Venus, Titan, icy moons and planetesimals, and discuss the implications for the enantioselective mutagenesis proposed as to be the driver of homochiralization. Earth is unusual in that spin-polarized muons dominate the cosmic radiation at its surface.  
\end{abstract}
\section{Introduction}
The origin of life  
is still a puzzle. However, it is a {\it chiral} puzzle: living organisms use only one class of enantiomer (molecule that has a mirror image) 
in the construction of proteins and nucleic acids, a fundamental property of life known as {\it homochirality} %\citep{mason1984,bonner1995}. 
\citep[see][for a recent review]{avnir2020}.
Homochirality allows biopolymers to adopt stable helical structures \citep{bada1995,nanda2007}, a natural conformation for biological macromolecules to adopt and one that plays a key role in fundamental biological processes \citep{totsingan2012}. While helical biopolymers have not been found, yet, in extraterrestrial environments, enantiomeric excesses (e.e.) of a few percent have been reported in the amino acids content of some meteorites \citep{pizzarello2006,burton2018}, but their relation to biological homochirality remains a mystery.

An important clue pertaining to the emergence of homochirality comes from studies of low energy, spin-polarized electrons which can promote enantioselective chemistry \citep[e.g.,][]{dreiling2014,rosenberg2019}. In addition, the Chirality-Induced Spin Selectivity, CISS, effect shows that {\it helical} biopolymers, specifically double strand DNA and $\alpha$-helical peptide, can act as spin filters for sub-eV electrons \citep{naaman2012}. This may be related to biological function and may play a role in the emergence of homochirality. 

 Cosmic rays, play a major role in generating mutation and promoting evolution on Earth. They create pions in the upper atmosphere which decay into highly polarized muons which comprise the most prevalent high energy particles at ground level. Muons decay into electrons and positrons which are also polarized. In \cite{globus2020}, henceforth GB20, it was proposed that this polarization could impose a small chiral preference on the development of one set of biological enantiomers over the other which could evolve to homochirality over millions or even billions of generations. Some mechanisms for expressing this preference were discussed, supplying a causal connection from a fundamental physics asymmetry to an equally fundamental asymmetry in biology.

There are three stages to this scheme: I. Understanding cosmic ray polarization in different environments, II. Estimating the chiral preference present in the mutation rate with a given irradiation and III. Translating this into an influence on biological evolution.  In GB20, the focus was on stage II and some semi-classical models of the interaction between cosmic rays and a chiral distribution of electrons in simple biopolymers were discussed.  A quantity called lodacity, ${\cal L}=\overline{\hat{\boldsymbol{\muup}}\cdot\hat{\textbf v}}$, was introduced because it is the cosmic ray magnetic moment, $\boldsymbol{\muup}$, not the spin, that mediates this interaction. The effects were small; larger effects may be exhibited in a quantum mechanical description. However it was also argued that the effects could have been sufficient to induce a enantioselective biological response and establish homochirality during stage III, especially if enantiomeric conflict is present.

In this paper, we focus on Stage I. We are most concerned with cosmic rays on Earth today. However, there are many quite different environments present within the solar system and evidence for life could be found here too. Or not. Either way, it is of considerable interest  to understand the evolution of cosmic ray polarization under different circumstances and so we also consider Mars, Titan, Venus, icy moons, and smaller bodies as well as Earth.

In Section~\ref{sec:asymmetry}, we explain the origin of the asymmetry in cosmic radiation. In Section~\ref{sec:depol}, we discuss the depolarization effects for muons and electrons in water, before presenting the different environments we consider in  Section~\ref{sec:founts}, and the doses in section~\ref{sec:doses}. We  discuss the biological implications in Section~\ref{sec:discussion}.

\section{Origin of polarization asymmetry in secondary cosmic radiation}\label{sec:asymmetry}
 Before discussing the results of detailed shower simulations, it is helpful to outline the development of the cosmic rays and their polarization from relativistic to the atomic energies that may be of most interest biologically. 
The cosmic rays observed in the solar system predominantly consist of hydrogen and ionized, typically stable nuclear isotopes (see the \citet{2020PTEP.2020h3C01P} and \citet{Gaisser:2016cr} for reviews). If the total flux of cosmic ray nuclei is converted into a flux of nucleons, one finds that only 15\% are neutrons and the rest protons. This charge asymmetry is conserved in the fluxes of secondary hadrons once protons interact with matter, since on average more $\pi^+$ and $K^+$ mesons are produced in the projectile fragmentation region. Hence, particles that emerge from hadronic processes and reach a ``fount'' (which we define as a location where trans-biotic polymers can develop, to distinguish it from a ``habitat'' where living organisms are well-established) 
will carry a net positive charge whereas particles from electromagnetic cascades have a charge ratio of one above the $\gamma \to \text{e}^+\text{e}^-$ threshold. At Earth, today, most particles at ground level are muons, which are tertiary cosmic rays predominantly produced through two-body decays of charged pions and kaons, {\it e.g.} $\pi^+ \to \mu^+\nu_\mu$. Since neutrinos are effectively massless, for our purposes, and  produced with left-handed helicity and in the pion's rest frame its zero spin is conserved, the muon is produced fully polarized with the opposite helicity of the neutrino. These two factors, charge asymmetry and parity violation of the weak interaction, imply that helicity state and charge have to be treated separately, resulting in four muon components $\mu^+_L$, $\mu^+_R$, $\mu^-_L$ and $\mu^-_R$ \citep{1965ICRC....2.1039V,1986ApJ...307...47D,lipari1993}, 
where $L$ ($R$) label correspond to negative (positive) helicity. Following \citep{lipari1993}, the minimal energy fraction the muon receives in the pion's rest frame is $r_\pi = (m_\mu/m_\pi)^2 \approx 0.57$, when it is emitted against the direction of movement, or $1$ when it coincides with the pion's direction. From this follows that the muon on average receives $\langle x \rangle = 1 - r_\pi/2 \approx 0.78$ of the pion's energy. The probability to for the helicity of muons when the pions are relativistic is
\begin{equation}
    P_R(x) = \frac{1}{1-r_\pi}\left(1-\frac{r_\pi}{x}\right),~~~
    P_L(x) = \frac{1}{1-r_\pi}\left(\frac{r_\pi}{x} - r_\pi\right),
\end{equation}
which gives roughly $2/3$ right-handed and the rest left-handed muons evaluated at $\langle x \rangle$. For the heavier charged kaons, 95\% of muons are right-handed what can be verified by replacing $m_\pi$ with $m_{{\rm K}^\pm}$.
Since we are only interested in the magnetic moment polarization, ${\hat{\boldsymbol{\muup}}\cdot\hat{\textbf v}}$, we denote by 
${\upharpoonleft \! \upharpoonright}$ as the sum of doses deposited by muons with $\mu^+_L + \mu^-_R$ (i.e. $\hat{\boldsymbol{\muup}}\cdot\hat{\textbf v}<0$) and ${\upharpoonleft \! \downharpoonright}$ as the sum of doses deposited by $\mu^+_R + \mu^-_L$ (i.e. $\hat{\boldsymbol{\muup}}\cdot\hat{\textbf v}>0$).The muons decay primarily into positrons and electrons together with two additional neutrinos. These also retain a significant spin polarization and lodacity ${\cal L}$, which we define to be the average of ${\hat{\boldsymbol{\muup}}\cdot\hat{\textbf v}}$ and is always of the same sign (GB20).

\begin{figure*}
\centering
\includegraphics[width=2\columnwidth]{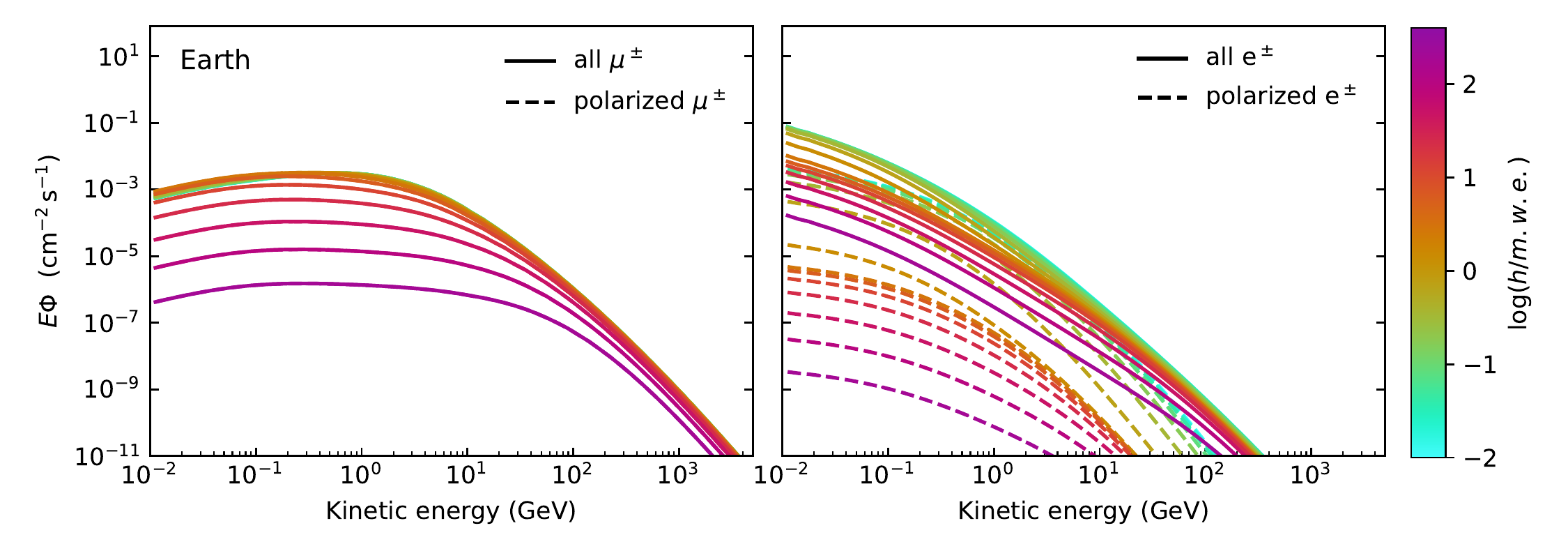}
\includegraphics[width=2\columnwidth]{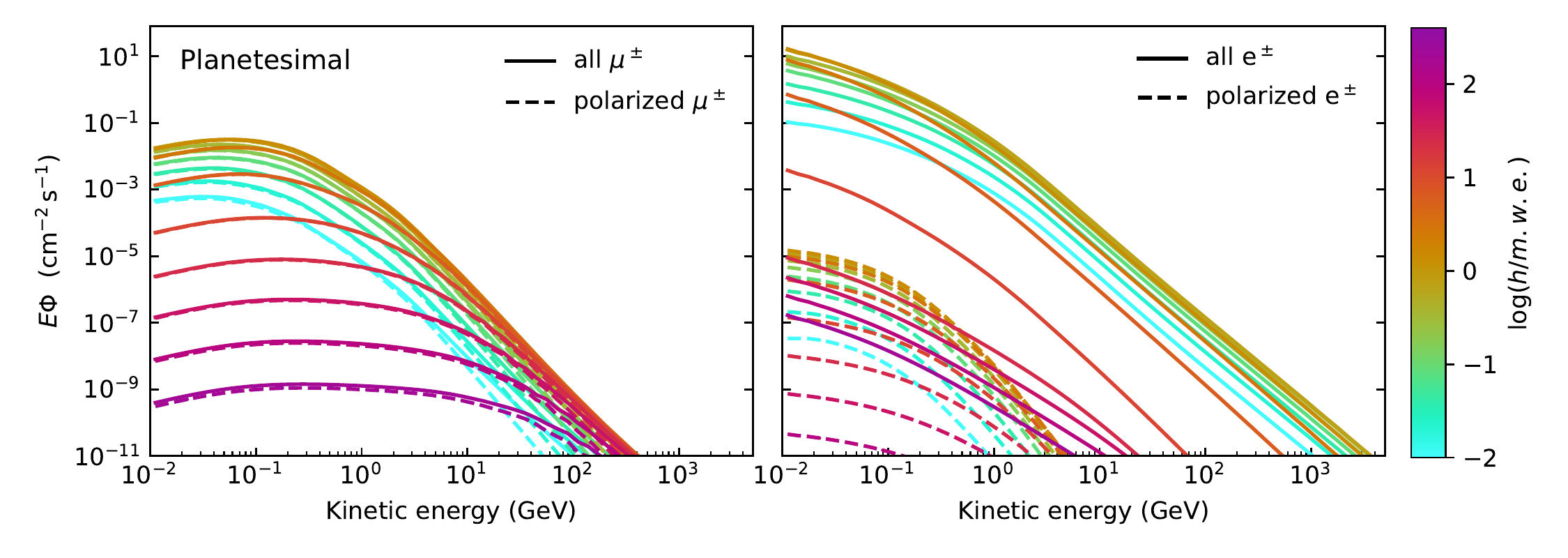}
\caption{Underground fluxes of muons and electrons on Earth (top) and on an icy moon without an atmosphere (bottom). The spectrum and composition of the primary cosmic rays are the one observed at Earth at the present epoch. 
In the left panels, the total $\mu^+ + \mu^-$ spectra at different depths are figured by the solid lines; the polarized muons by dashed lines. The fraction of polarized muons is close to unity since they mostly come from pion decay. In the right panels, the total $e^+ + e^-$ spectra, i.e. those produced by all lepto-hadronic processes and also from bremsstrahlung pairs, are shown by solid lines at different depths. The polarized $e^+ + e^-$ from decay of muons are shown by dashed lines. %.(This is for the large body case. Small body case is given in appendix.)
}
\label{fig:mu_el_water}
\end{figure*}

\section{Depolarization Effects}\label{sec:depol}
\subsection{Muons}
%A spectrum of cosmic rays is incident on the atmosphere over a hemisphere. 
%Most of the hadrons are stopped at {\bf a depth of $\sim$~100~${\rm g\,cm}^{-2}$},
 Hadronic cascades  produce muons which are subject to ionization loss and decay into electrons and positrons.  With a vertical atmospheric grammage $\sim1000\,{\rm g\,cm}^{-2}$ and a scale height $\sim10\,{\rm km}$, the surface muon particle flux %, per unit $\ln T$, 
is broad in energy but peaks at a few GeV. Lower energy muons mostly lose energy and decay before reaching the ground. Fewer high energy muons are produced and they are subject to extra radiative and pair losses. 

For illustration, we consider underwater fate of few GeV muons and this elucidates what happens in other environments. Muons lose energy in water though ionization loss on a timescale 
%\begin{eqnarray}
%t_{\rm loss,\,\mu}\equiv|dt/d\ln T|&\sim&20\,T_{\rm G}^{0.9}\,{\rm ns};\,0.1\lesssim T_{\rm G}\lesssim100,\nonumber\\
%&\sim&8\, T_{\rm M}^{1.2}\,{\rm ps};\,0.01\lesssim T_{\rm M}\lesssim100,\nonumber\\
%&\sim&35\,{\rm fs};\,0.1\lesssim T_{\rm k}\lesssim10.
%\end{eqnarray}
\begin{eqnarray}
t_{\rm loss}^\mu\equiv|dt/d\ln T|&\sim&20\,T_{\rm G}^{0.9}\,{\rm ns};\,0.3\lesssim T_{\rm G}\lesssim100,\nonumber\\
&\sim&6\, T_{\rm M}^{1.2}\,{\rm ps};\,0.02\lesssim T_{\rm M}\lesssim300,\nonumber\\
&\sim&60\,{\rm fs};\,0.1\lesssim T_{\rm k}\lesssim20\,,
\end{eqnarray}
where the kinetic energy $T_{\rm G,\,M,\,k}\equiv T/({\rm 1\,GeV,\,1\,MeV,\,1\,keV})$. Muon decay can be ignored on these timescales and once they become subrelativistic, muon ranges decrease $\propto T$ and so they essentially stop on the spot.

We show in Appendix~\ref{appendix:E_loss} that, for a particle that slows through ionization loss, the muon scattering is relatively unimportant. This implies that muons continue in the same direction and that they remain polarized because their spins are much less deflected in encounters than their velocities. 
The muon lodacity will decrease according to ${\cal L}\propto\exp(-K\lambda)$, where the ``lethargy'' $\lambda={\rm sech}^{-1}(v/c)$ and the depolarization factor $K=0.037$ for muons in water (see Eq.~\ref{eq:lethargy} in Appendix~\ref{appendix:E_loss}). This implies a lodacity reduction by a factor of only  $\sim0.83$ as the muons cool to a kinetic energy $T_{\rm k}\sim10$.\footnote{The degree of depolarization due to ionization during slowing-down can also be estimated using a formula given by \citet{akylas1977}; they find that depolarization is negligible.} 

Down to few keV, the depolarization is ignorable. Below a few keV, there is a significant difference between positive and negative muons. Positive muons capture electrons with atomic cross sections \citep[e.g.][]{kulhar2004} when their speeds are comparable with those of valence electrons, $T_{\rm k}\sim10$, to form muonium, Mu. This reduces the muon polarization by by a half \citep{nagamine2003}. There will be further scatterings and exchanges, leading to further depolarization as the Mu thermalizes. 
%, which may undergo exchanges before the muon decays in $\sim 2\,\mu{\rm s}$. Once the positive muon is coupled with an unpolarized electron to form Mu, the polarization of Mu isimmediately reduced to 50\% \citep{nagamine2003} and will be thus further reduced by thermalization and subsequent scatterings.

Negative muons, by contrast, continue to cool through atomic interactions mostly through excitation as they are too slow to ionize valence electrons. However, their deflections remain small and they retain their polarization until they reach atomic energies, $T_{\rm k}\sim0.03$. At this point,  they form muonic atoms and rapidly cascade into high energy ground states very close to the host nuclei before decaying. The capture cross section is spin-dependent and may be sensitive to the chirality $\cal M$ of the molecule (GB20).  Modest depolarization will occur within the atom through spin-orbit interaction. For instance, negative muons stopped in carbon retain about 15\% of their initial polarization in the K-shell of spin-zero nuclei before they decay \citep{garwin1957}. The de-excitation cascade  to the ground state is accompanied by the emission of Auger electrons and circularly polarized X-rays which can also transfer some polarization.

To summarize, muons  retain almost complete polarization until they form Mu or muonic atoms. They transport polarization efficiently from high to low energies where they can have a biological impact. This is not the case for the electrons, as we discuss in the next section.

\subsection{Electrons}
In Fig.~\ref{fig:mu_el_water}, we show the underground fluxes of muons and the accompanying electrons as a function of depth (in meter water equivalent), on Earth and on a moon without an atmosphere. While the muons are highly polarized, the electron polarization is negligible. The basic reason is that electrons scatter faster than they cool (Appendix B). This implies that polarized electrons made in muon decays in the atmosphere are rapidly depolarized and do not reach the surface. Electrons produced in muon decays below the surface are also unimportant.

However, the muons cool by creating unpolarized, knock-on electrons, an unavoidable ``entourage'' of high energy electrons that must accompany the muons.  
 High-energy muons ($T >> {\rm GeV}$) that dominate at larger depths emit photons through bremsstrahlung $\mu^\pm \to \mu^\pm + \gamma$ and initiate electromagnetic sub-cascades through $\gamma \to \text{e}^+ + \text{e}^-$. These contribute to the ``entourage'' of unpolarized electrons, hence diluting effects induced by the high-energy muon polarization. A given biological molecule will evolve in the presence of a spectrum of both particles. While we do not understand their separate effects, a first step is to estimate their relative densities. This is discussed more in Appendices \ref{appendix:electrons} and \ref{appendix:codes}.

\begin{figure*}[!h]
\centering
\includegraphics[width=.9\textwidth]{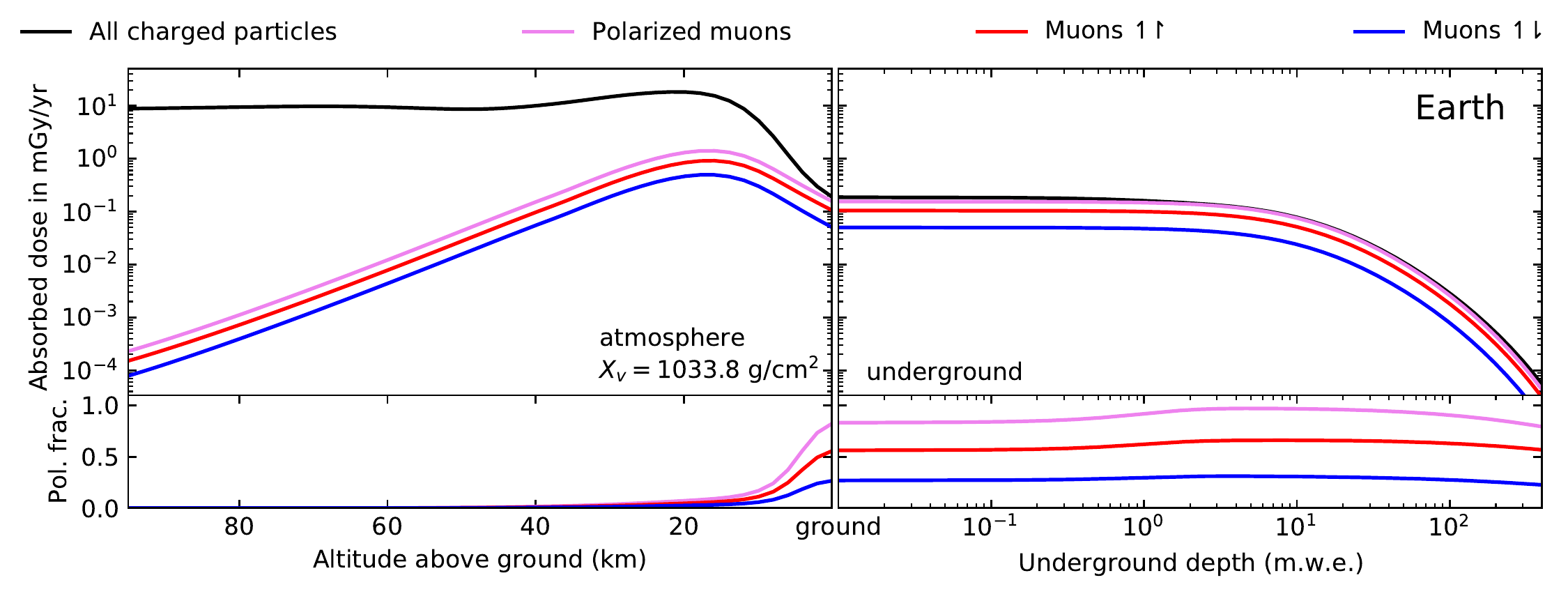}
\includegraphics[width=.9\textwidth]{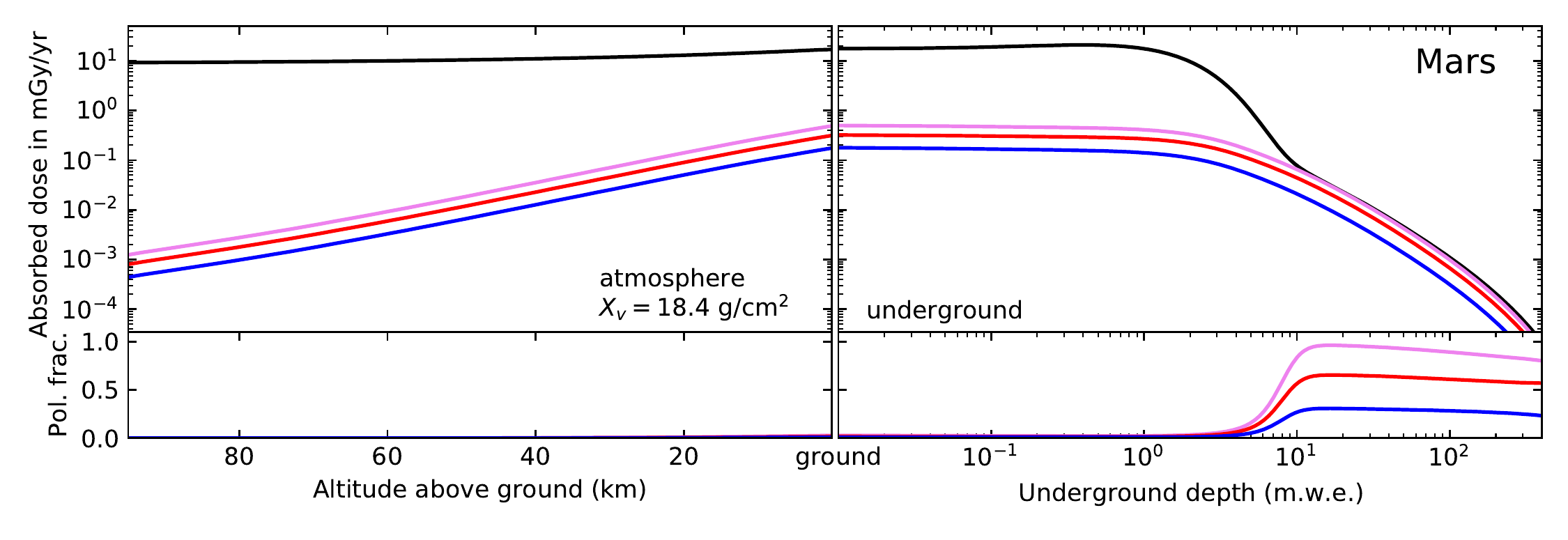}
\includegraphics[width=.9\textwidth]{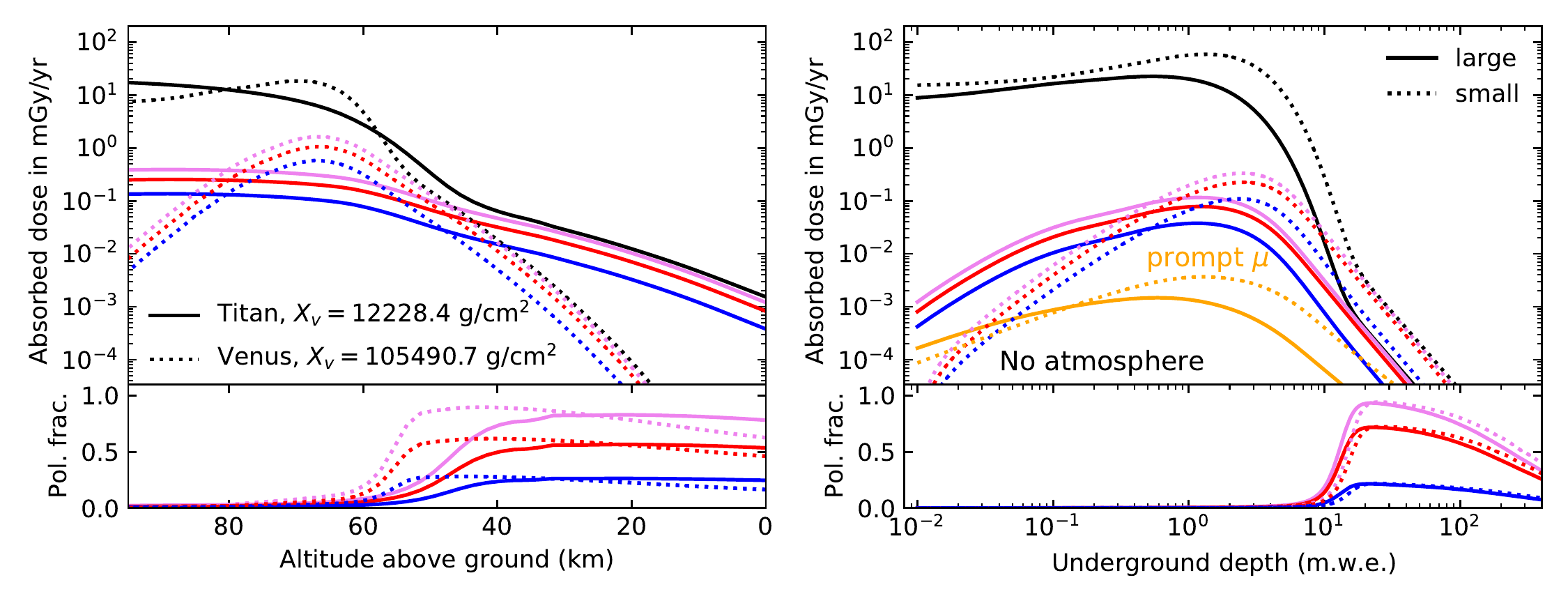}
\caption{Top and Middle panels: Absorbed dose rate due to cosmic ray radiation at Earth and Mars. The black curves are obtained by summing the ionization losses from all charged particles. Red curves (${\upharpoonleft \! \upharpoonright}$) are the sum of doses deposited by muons with $\mu^+_L + \mu^-_R$ whereas the blue (${\upharpoonleft \! \downharpoonright}$) are those with $\mu^+_R + \mu^-_L$. The purple curve is the total muon component.  
At Earth polarized muons constitute the dominant source of cosmic radiation at sea level and shallow underground depths. On Mars, polarized muons become the main source of cosmic radiation at $\sim 10$ meters water equivalent (m.w.e). To estimate the underground fluxes in rock, the x-axis of the right-hand side panels can be multiplied with $\rho_\text{rock}/({\rm g/cm}^3)$. 
Bottom left panel: Same calculation for the dense atmospheres of Titan and Venus.
Bottom right panel: Same calculation for planetesimals, like Earth's moon, or asteroids, protoplanets or comets, with a negligible atmosphere. ``Large'' signifies bodies with size much larger than the depth for which up-going particles can be neglected. In ``small'' bodies, the radiation deposited is roughly isotropic and shown for the center of a sphere with radius equal to the depth. The additional unpolarized, prompt muons, shown in orange, originate from three-body decays of short-lived charmed and unflavored mesons and from electromagnetic pair production (see Figure~\ref{fig:deep_underground} for more detail). Note that the dose is plotted logarithmically for clarity.
}
\label{fig:polarized_depth}
\end{figure*}

\section{Founts}\label{sec:founts}
Now we have explored the effect of depolarization and dilution, we will calculate the polarized radiation doses at various environments: Mars, Titan, Venus, and  bodies without an atmosphere (moons, asteroids). We are interested in {\it founts}, that we defined in the introduction as places where trans-biotic polymers can develop. The search for life in our solar system has been focused on finding water, one of the key prerequisites for life as we know it, in the form of subsurface lakes or oceans. In some environments,  geologic and hydrologic activity could have enabled physical processes necessary for the growth of the first {\it helical} biopolymers \citep[see][for a review]{camprubi2019}.

\subsection{Worlds with a thin or negligible atmosphere}

Mars is a frozen world punctuated by periods of hydrologic activity \citep{cabrol2018}. Recently, a 20-km-wide lake of liquid water has been detected underneath  solid ice in the Planum Australe region, at approximately 1.5~kilometer depth \citep[][]{orosei2018}. Although there is still no evidence for life on Mars, the red planet has once deployed environments characteristics of  wet-dry cycling, a process thought to have enabled the production of the first helical biopolymers \citep{ross2016}; the Mars Exploration Rover Spirit reported  3.65-billion-year-old hot spring deposits, approximatively of the same age as the Dresser hot springs on Earth that showed evidence for microbial life \citep{van2017, damer2020}. If microbial life started in similar hot springs, it is likely that after Mars' geological death and the loss of its atmosphere, microbial life would have only be able to survive underground, since Martian subsurface  seems to have the key ingredients to support life for hundreds of millions of years \citep{michalski2018, checinska2019}. 

Life on Mars could have originated when its atmosphere was much denser than today and comparable to the current Earth's atmosphere. In consequence, the radiation dose and spectrum penetrating to the surface are likely to be quite different for Noachian and present-day Mars as discussed by \citet{lingam2018}. Mars magnetic field could have been larger in the Noachian \citep{mittelholz2018} but this would not affect the muon polarization (GB20).  As for the possible (intermittent) variations in the cosmic ray flux, we consider them in Section~\ref{sec:doses}.

Evidence has  accumulated that subsurface liquid regions (lakes or oceans) exist beneath the surface of the icy moon of Jupiter, Europa, and the icy moon of Saturn, Enceladus \citep{thomas2016}. Mass spectrometers aboard the Cassini spacecraft found that the plumes of Enceladus contain water, salts, ammonia, carbon dioxide, and small and large organic molecules, suggesting that Enceladus' ocean may entertain an organic chemistry \citep{khawaja2019}. In these environments, the atmosphere is negligible or absent (as in asteroids, comets or meteroids). Like on Mars, we will see that in these worlds where the atmosphere is absent, the spin-polarization transport is different than on Earth.

\subsection{Worlds with a very dense atmosphere}

We also consider worlds with atmospheres denser than the present Earth. Titan, Enceladus and Europa are among the most geologically active worlds in our solar system but in contrast to Enceladus and Europa, Titan possess a very dense atmosphere. The essential chemical building blocks for life are present in Titan's atmosphere \citep{yung1984}. Titan is also thought to have a subsurface ocean of water \citep{iess2012}\footnote{It is possible that major impacts on a moon like Titan, Europa or Enceladus could expel large, icy fragments with liquid, life-containing interiors (or even surface oceans) into interplanetary space. Absent a heating source, the water would soon freeze, fossilizing its contents. It is surely of long-term interest to locate, visit and examine such bodies \citep{dyson1997}.}. For Venus, the habitable region lies  in the upper atmosphere \citep{morowitz1967}. %Recently, there has been a lot of debate on the possible detection of %phosphine and also  of the simplest achiral  amino acid, glycine, in the atmosphere of Venus but these results still need confirmation %\citep{greaves2020,snellen2020, manna2020}. \citep{manna2020}.

\section{Polarized radiation doses}\label{sec:doses}
The current absorbed dose rates from cosmic radiation in different environments, are presented in Figure \ref{fig:polarized_depth}. It is an interesting coincidence that for Earth, at ground, 85\% of ionizing cosmic radiation is due to the polarized muonic component that increases to almost 100\% two meter water equivalent (m.w.e.) below the surface. 

In other bodies of the solar system, the presence or absence of atmosphere leads to different results. For Mars (middle panel of Fig.~\ref{fig:polarized_depth}), cosmic rays are only slightly attenuated by the thin Martian atmosphere and the  polarized muons become the dominant source of cosmic radiation at $\sim10$ m.w.e. with a dose of $\sim 0.01$ mGy/yr, much lower than the $0.2$ mGy/yr on Earth. For planetesimals without an atmosphere or smaller bodies of the solar system (bottom right panel of Fig.~\ref{fig:polarized_depth}), the polarized muons component dominates at $\sim15$ m.w.e. and with only one tenth of the intensity.  In a dense, solid medium, the interaction length is much shorter than the decay length, leading predominantly to (re-)interaction of unstable hadrons rather than their decay. This suppresses the production of high energy muons and leads a swiftly decreasing polarized muon component as a function of depth. 

In Titan and Venus, %, hadronic cascades peak at a depth of $\sim100\,{\rm g\,cm}^{-2}$, or altitude $\sim40-50\,{\rm km}$ and 
polarized muons dominate the radiation at altitudes $40$, $50\,{\rm km}$. The surface irradiation, comparable to that below $\sim400\,{\rm m}$ of rock is negligible.

\begin{figure}
\centering
\includegraphics[width=\columnwidth]{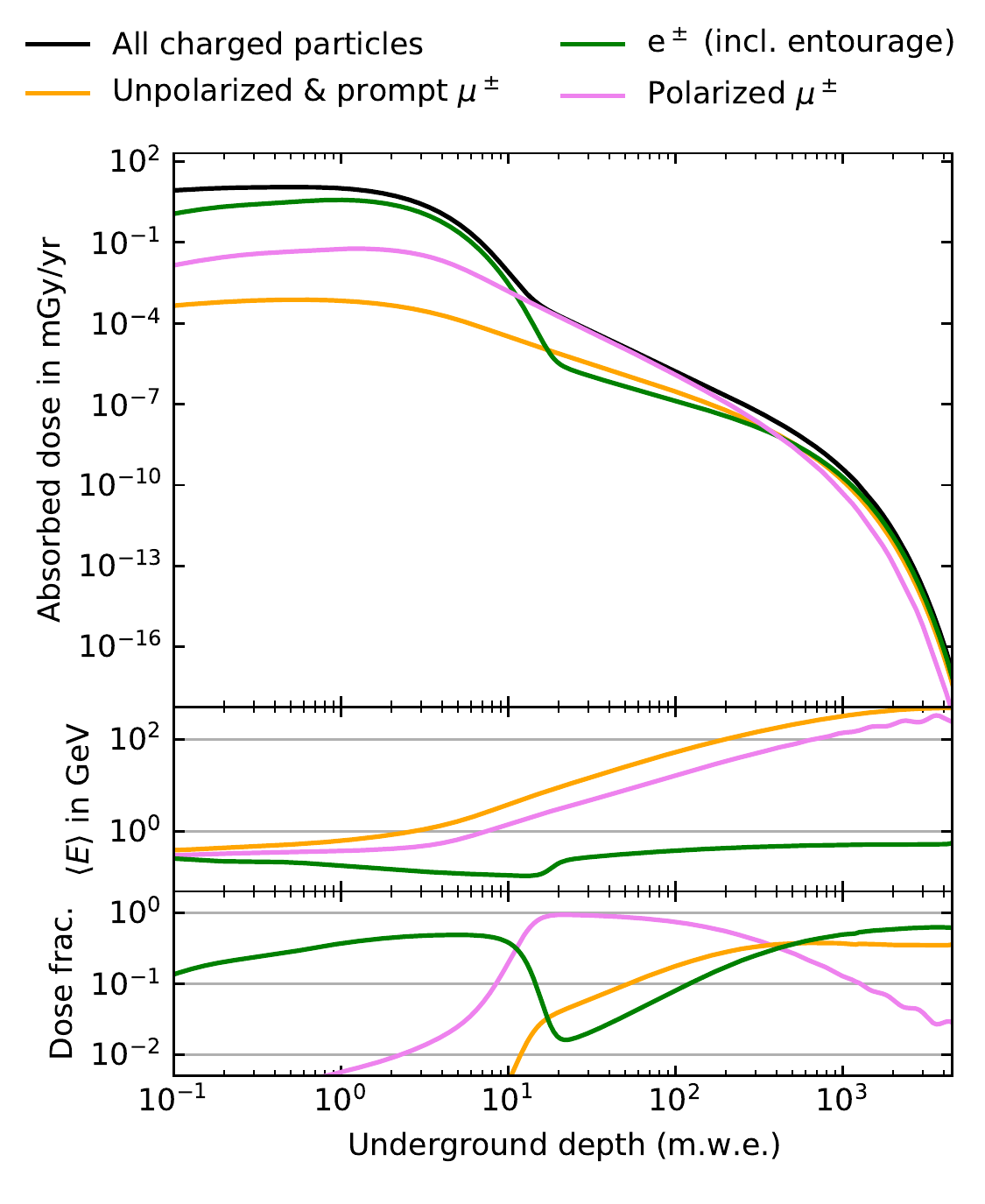}
\caption{Dilution of polarized radiation by electronic entourage at deep undergrounds. This figure extends the case of a "large" planetesimal from the bottom right panel of Figure~\ref{fig:polarized_depth}. The depths of several km.w.e. are more representative environments found in (for example) Enceladus' crust. At greater depths low energy muons have stopped or decayed, and those which survive have increasingly higher $\langle E_\mu \rangle \gg$ GeV, as illustrated in middle panel. High-energy muons lose energy predominantly by emitting high-energy bremsstrahlung rather than through ionization. These gamma rays convert into pairs that join the electronic ``entourage''.}
\label{fig:deep_underground}
\end{figure}

Figure~\ref{fig:deep_underground} demonstrates that the dose deposition in a large icy body at km.w.e.~depths is many orders of magnitude feebler compared to the scenarios that we study at a few m.w.e. The fraction of the polarised dose drops due to two effects: 1) kilometer depths are reached predominantly by the higher-energy, unpolarized, prompt muons \citep{2019PhRvD.100j3018F}, {\it i.e.}~from decays of unflavored mesons and heavy charmed hadrons, or from electromagnetic pair production; 2) the average muon energy grows because low energy muons stop or decay at shallower depths. The stopping power of high-energy muons become increasingly radiative, which deposits energy into the unpolarized electromagnetic entourage via bremsstrahlung.

\begin{figure*}
\centering
\includegraphics[width=\textwidth]{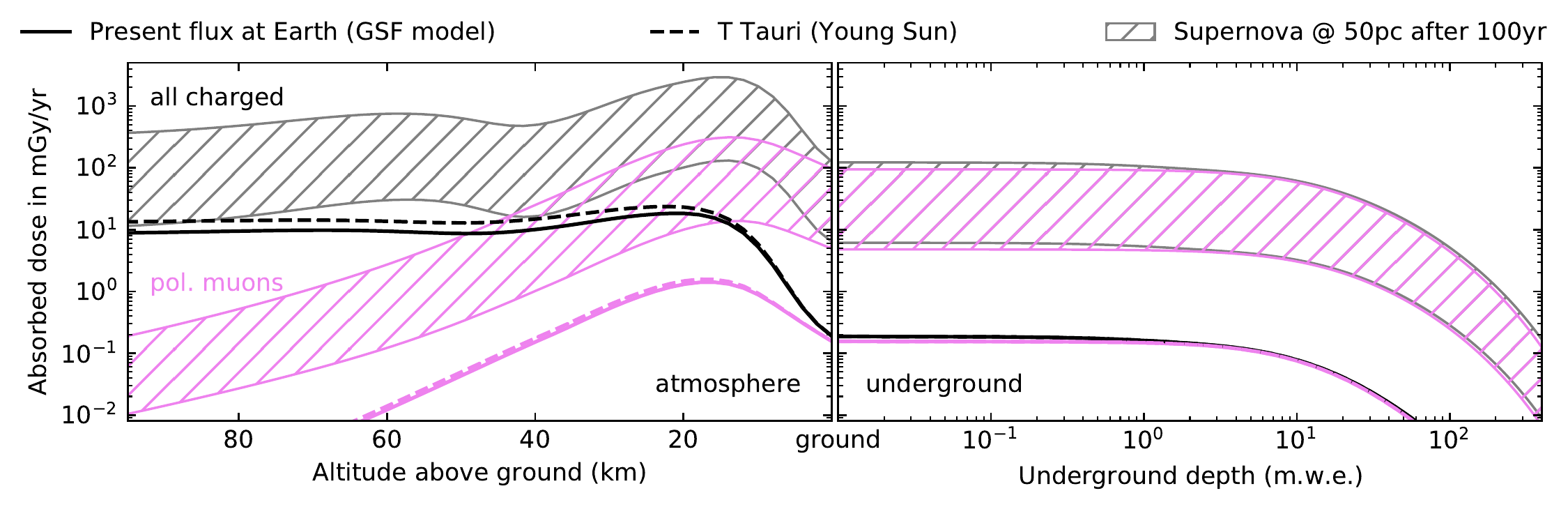}
\caption{Absorbed dose at Earth for the presently observed cosmic ray flux according to the GSF model \citep{2017ICRC...35..533D} as in Fig.~\ref{fig:polarized_depth}, a phase in which the young Sun is modeled with a T Tauri star spectrum from Fig.~10 in \citet{2018A&A...614A.111P}, and for a nearby supernova 100 years after explosion at 50pc \citep{2017ApJ...840..105M}. Despite the high proton densities assumed in the Taurosphere, the impact on the underground fluxes is negligible since most protons have energies below the particle production threshold and hence do not produce additional muons. In the nearby supernova scenario the upper edge represents case A from \citet{2017ApJ...840..105M} and the lower case B, respectively. Shown is the maximal fluxes expected 100 years after explosion that drop significantly within a few kyr.}
\label{fig:crflux_dose}
\end{figure*}

The terrestrial polarized muonic dose has a maximum of $\sim 0.2$ mGy/yr at the current epoch at ground level, but could have been larger during the Hadean period where a young sun could have accelerated cosmic rays beyond $\sim10$ GeV. If the cutoff of stellar cosmic rays arises below that energy (Stellar proton events (SPEs) or a young T Tauri cosmic ray spectrum as modeled by \citet{2018A&A...614A.111P}, this would only change the radiation dose by a factor $\sim2$, as we show in Fig.~\ref{fig:crflux_dose}. % {\bf  \citep[See also][for the gamma-ray burst scenario]{atri2014}.
 All environments harbor other natural sources of radiation that may exceed the dose deposited by cosmic rays. About 3.5  billion years ago, life was likely exposed to a background radiation field of $\sim6$ mGy/yr \citep{karam1999}. 
However, more energetic transient events such a nearby  supernova could increase the doses due to cosmic muons  up to $\sim100$ mGy/yr at ground level.%  \citep[See also][for the gamma-ray burst scenario]{atri2014}.
We show in Fig.~\ref{fig:crflux_dose} the absorbed dose at Earth 100yrs after the explosion of a nearby supernova \citep[][]{2017ApJ...840..105M} compared to the actual doses.  Since the muons retain their lodacity, higher radiation doses from cosmic muons could have altered enantioselectively the evolutionary trajectories in shorter timescales due to higher mutation rates.
The closest supernova explosion over the age of the Earth was roughly $\sim5-10$~pc away \citep[e.g.][]{ellis1995}. The prompt electromagnetic (UV, X-ray, $\gamma$-ray) radiation could have had serious biological effects but would not be significantly chiral. By contrast, the pulse of freshly-accelerated cosmic rays is chiral and should have delivered a fluence $\sim10-100\,{\rm MJ\,m}^{-2}$ over several thousand years, overwhelming the defenses presented by the solar wind, the magnetosphere and the atmosphere. For comparison, this is $\sim10^4-10^5$ times the fluence associated with a giant, "Carrington"-strength solar flare, albeit delivered over a much longer timescale. Activity associated with the nucleus of our Galaxy, including tidal disruption events \citep{pacetti2020}, might also be biologically consequential. It is therefore possible, though far from assured, that life was initiated, not terminated, by such cataclysmic events.

%\noteAF{All environments harbor other natural sources of radiation that may exceed the dose deposited by cosmic rays, diluting their chiral effect. [Comment: I think, this implies a minimal age of the planetary crust to shield an environment from the exchange of radioactive isotopes with the mantle, such as Thorium or Uranium ($\to$Radon).}

\section{Discussion}\label{sec:discussion}

%We showed in this {\it Letter} that while on Earth, the highest dose of spin-polarized muons  is $\sim$0.2 mGy/yr, with a dose almost constant from zero to 10 m.w.e. underground, in other environments, the polarized muons dominates $\sim$ 40km above the ground for Venus and titan, or more than 10 meters below the surface for bodies with thin or negligible atmosphere such as moons, asteroids or comets. Muons  retain almost a complete polarization until they form Mu or muonic atoms. They   are therefore very efficient to transport the polarization down to energies where they can have a biological impact.

So far, scenarios for the emergence of biological homochirality involved two processes. The first one is the generation of an initial enantiomeric excess (e.e.) in biologically relevant but still simple chiral monomers, such as amino acids or sugars. Various ways to achieve an initial e.e. have been proposed:  parity-violating energy differences at the molecular level \citep[][and references therein]{szabo1999}, asymmetric destruction due to irradiation by beta particles in radioactive decay \citep{vester1959}, differential absorption of circularly polarized light \citep{hendecourt2019}, asymmetric adsorption on chiral mineral surfaces like clays \citep{hazen2001}. Because the resulting enantiomeric excesses are very small \citep[only up to few percent, see review by][]{burton2018}, a second necessary process is the amplification of the initial e.e. from low to high values. Enantioselective chemistry might be able to amplify the e.e. up to 99.5\% \citep{soai2000} but the validity of such chemical processes in natural environments is still debated. 

In contrast, the scenario proposed by GB20  only involves enantioselective mutagenesis, for which the timescale depends on the mutation rates of the first trans-biotic polymers and the magnitude of the  bias.  
 In order for the particular enantiomeric choice observed in living systems on Earth to be causally related to the choice expressed by the weak interaction and mediated by cosmic rays, we argued in GB20 that there needs to be a coupling between the lodacity, $\cal L$, of the cosmic rays and the molecular chirality, $\cal M$, of the molecular structure. The mutation rate has to include a term proportional to the product of these two pseudoscalars. The possibilities include an electromagnetic interaction in which the scalar product of the electric and magnetic dipoles or, equivalently ${\bf E}\cdot{\bf B}$ had a preferred sign. There is no good reason for this to occur in an isolated atom. However, it is just what is expected in a segment of a helical molecule containing a conduction or a magnetization current. In this case  $<{\bf E}\cdot{\bf B}>$ is negative (positive) for a right- (left-) handed helix, independent of the polarity.

We conclude with two thoughts. Firstly, Earth is the only body in the solar system with an atmosphere just the right density and scale height to ensure that muons dominate the cosmic radiation at ground level and for $\sim10\,{\rm m.w.e.}$ underground. It is thick enough to stop the other constituents, thin enough for the muons to survive.  For worlds with thin or negligible atmosphere, such as Mars, polarized muons are the dominant source of cosmic radiation at depths of 10 m.w.e. while for worlds with dense atmospheres, like Venus or Titan, muons dominate below $\sim$40 km altitude but the radiation doses are negligible at the surface. Now, muons are the one constituent of cosmic radiation that is created with and retains high polarization as it decelerates to speeds $\lesssim\alpha c$ when it can promote mutation of trans-biotic biopolymers. Might this be connected with the fact that Earth is the only body known to harbor life?
However, we know that Mars, Venus, and Earth have undergone significant evolution. Early in the Solar system's history, Mars probably had an atmosphere of ~0.5-1 bar \citep{lingam2018} and Venus may have possessed an atmosphere of $\sim$1 bar until a Gyr ago or so \citep[e.g.][]{way2016}. However, if life originated in the early solar system, conditions could have been different from those we experience today. Therefore the special character of Earth's atmosphere might be more relevant to life's development than to its origin.

The second thought is that the relative importance of the muons and their electronic entourage for mutation is not understood. In GB20, ionization was taken as a proxy for mutation. If this is the case, then collisional ionization by electrons is probably more frequent than muon capture, diluting the chiral effect of the cosmic rays.  Muons have advantages of better depth-dose distributions.   %The Bragg peak of muons is sharper compared to that of \noteRDB{Highly relevant to radiotherapy but I am not sure that it is for cosmic rays.} electrons \citep[e.g.][]{mokhov1999} and therefore, in a living organism, muons can deposit less dose in the protective membranes and  most dose in the target DNA.
However, even ionization may be reparable and not so conducive to genetic alteration.   There could be an appreciable chiral contribution to the Hamiltonian for the formation of a muonic atom or Mu in a helical molecule. The fact that negative muons retain their polarization down to the energies where they can undergo a spin dependent capture, may also have an effect on enantioselective mutagenesis. A single muonic atom decay will produce polarized X-rays and Michel electrons\footnote{Michel electrons have a well known energy distribution with a sharp peak at 52.83 MeV and an asymmetric angular distribution if the parent muon is polarized, and would contribute to the chiral bias.} and will surely fracture the molecule.   
Even the pickup of an electron in Mu formation could be quite disruptive. 
Perhaps it is muons, not electrons, that dominate mutation of trans-biotic biopolymers, thereby increasing the chiral transfer and accelerating evolution to homochirality.

These matters and the prospect of performing experimental investigations will be discussed in a future paper.

\section*{Acknowledgements}
We are thankful to David Deamer, Bruce Dunham, Akimichi Taketa,  for helpful discussions and encouragements. 
The research of NG was supported by New York University and the Simons Foundation.
AF completed his work as JSPS International Research Fellow (JSPS KAKENHI Grant Number 19F19750).

\bibliography{scibib}

\appendix

\section{Scattering and ionization loss}\label{appendix:E_loss}
The rate of loss of kinetic energy $T$ by a cosmic ray in an environment where the molecular density is $n$, is dominated by the ionization of electrons and can be approximated by 
\begin{equation}
\frac{dT}{dt}=-\frac{4\pi n}{m_ev}\left(\frac{e^2}{4\pi\epsilon_0}\right)^2\ln\Lambda\sum_iZ_i,
\label{electronic}
\end{equation}
where $v$ is the speed, the sum is over all the atomic numbers $Z_i$ of the nuclei in the molecule and $\Lambda$ is the ratio of the maximum to the minimum impact parameter given roughly by $2p^2/m_eI$, where $p$ is the momentum and $I$ is a mean ionization potential. As the energy loss is dominated by distant encounters, the energy fluctuations are small and the kinetic energy can be regarded as function of time. 

Nuclear recoils do not contribute to the stopping power but they dominate the scattering. Consider a cosmic ray with momentum {\bf p} making an angle $\theta$ with the direction of its magnetic moment, $\hat{\boldsymbol\muup}$ which we can regard as fixed in space. The mean square rate of change of the transverse momentum by nuclear and electron collisions can be similarly expressed as
\begin{equation}
\left<\frac{d<\Delta p^2>}{dt}\right>=\frac{8\pi n}v\left(\frac{e^2}{4\pi\epsilon_0}\right)^2\ln\Lambda\left(\sum_iZ_i+\sum_iZ_i^2\right).
\label{nuclear}
\end{equation}
Now, the lodacity, ${\cal L}\equiv<\cos\theta>=<\hat{\boldsymbol\muup}\cdot{\bf p}>/p$, and if we average over azimuth in the scattering, we find that $\Delta\ln<{\cal L}>=-<\Delta p^2>/2p^2$. Again the fluctuations will be small and can be ignored. 

It is helpful to introduce a quantity
\begin{equation}
\lambda=m\int_T^\infty\frac{dT'}{p^2}={\rm sech}^{-1}\frac vc=\ln\left(\frac{pc}T\right),
\end{equation}
which we call the ``lethargy'' for a cosmic ray with rest mass $m$. We can then combine these three equations to obtain
\begin{equation}
\frac{d\ln{\cal L}}{d\lambda}=-K=\frac{m_e}m\left(1+\frac{\sum_iZ_i^2}{\sum_iZ_i}\right),
\label{eq:lethargy}
\end{equation}
and so, ${\cal L}={\cal L}_0\exp(-K\lambda)={\cal L}_0(T/pc)^K$, where $K$ is the depolarization factor. For cosmic ray muons, electrons in water, $K\sim0.037,7.6$. 

\section{Knock-on electrons from cosmic ray muons under water}\label{appendix:electrons}
Consider, as a specific example, muons incident upon the surface of the ocean on Earth. We have argued that muons decelerate through ionization loss while retaining most of their initial lodacity. For the energies that dominate the particle flux at the surface of the earth, $\sim 1-10\,{\rm GeV}$, they stop at a depth of tens of meters before they decay. When  they become non-relativistic, their range scales $\propto T^2$. After they cool, they will be captured on atoms, ($\mu^-$), or capture an electron, ($\mu^+$). By contrast, electrons scatter before they decelerate. Polarized electrons created through muon decay are rapidly depolarized and are unimportant compared with the knock-on electrons which will cool and thermalize on the spot by creating more knock-on electrons, forming an unpolarized ``entourage'', accompanying the muons. 

The polarized muons and the unpolarized electrons will compete in creating mutations of trans-biotic molecules, diluting the chiral transfer. In order to quantify this effect, we first consider a fixed number density high energy muons crossing a volume that is large enough to allow the knock-on electrons to cool locally and small enough for the muons not to cool in traversing it. Introduce a conserved number current in energy space of high energy muons $I^\mu$, the number of particles cooling through a kinetic energy $T$ per unit volume and time. Let this be established at some moderately relativistic energy $T_{\rm CR}\sim3\,{\rm GeV}$. Next, approximate the rate per unit volume, $S^{\mu\,e}(T';T)$, at which a muon, with kinetic energy $T$, produces electrons with kinetic energy $T'<T$ lying in $dT'$. Now, $\int^TdT'T'S^{\mu\,e}(T';T)\sim T/t_{\rm loss}^\mu(T)$ c.f. Eq.~(1). Likewise we introduce the current of electrons in energy space  $I^e(T)$, with $\int^TdT'T'S_{e\,e}(T';T)\sim T/t_{\rm loss}^e(T)$. $I^e(T)$ satisfies
\begin{equation}
\frac{dI^e}{dT}+\int_T \frac{dT'}{T'}[S^{e\,e}(T;T')t_{\rm loss}^e(T')I^e(T')+S^{\mu\,e}(T;T')t_{\rm loss}^\mu(T')I^\mu]=0.
\label{eq:electrons_current}
\end{equation}

Next multiply Eq.~\ref{eq:electrons_current} by $T$ and integrate over $dT$. We find the approximate solution 
\begin{equation}
I^e(T)T\sim  I^\mu T_{\rm CR} 
\end{equation}

In other words, the knock-on electron energy current in energy space at low energy is roughly the same as the  energy current for the muons at high energy.

In this paper we are most interested in the muon and electrons spectra which can then contribute to the biological evolution of trans-biotic polymers and introduce a chiral bias. Many processes contribute to the full picture as we shall discuss elsewhere. For the moment, as an example, let us consider $\sim10\,{\rm keV}$ electrons which are known to induce mutation in DNA. These are produced by muons with energy $T\gtrsim1\,{\rm MeV}$. As the energy current in muons with energy $\sim3\,{\rm GeV}$ is roughly the same as the energy flowing through $\sim10\,{\rm keV}$ as electrons, we can conclude that the number current of $\sim10\,{\rm keV}$ electrons is roughly $3\,{\rm GeV}/10\,{\rm keV}\sim3\times10^5$ times the conserved number current of muons. The implications of this for biological processes then depend on combining these rates with the relevant cross-sections.

\section{Computation of particle fluxes}\label{appendix:codes}

We compute differential particle fluxes $\Phi(E,X)$, where $E$ is the kinetic energy and $X(h) = \int_h^\infty{\rm d}\ell~\rho(\ell)$ is the slant depth in g/cm$^2$, using the one-dimensional cascade equation solver \mceq{}\footnote{\url{https://github.com/afedynitch/MCEq} (Version 1.3)} \citep{2015EPJWC..9908001F} that is optimized for calculations of inclusive atmospheric lepton fluxes at high energies above a few GeV \citep{2019PhRvD.100j3018F}. We have extend the kinetic energy range down to 10 MeV and for computations in different media, such as carbon-dioxide and water by using hadronic cross sections and particle production tables from the {\sc DPMJET-III-19.1}\footnote{\url{https://github.com/afedynitch/DPMJET}} event generator \citep{2001amc..book.1033R}. Electromagnetic cross sections have been computed using the numerical routines from \citet{2019ICRC...36..961M}. Ionization losses $\langle {\rm d}E/{\rm d}X \rangle$ are based on tables from \citep{2020PTEP.2020h3C01P,ionization_tables} and are tracked for each charged particle whereas energy deposition by tertiary electromagnetic cascades from fast neutrons are neglected.  \mceq{} adapts the notation and formulae for muon polarization in the relativistic approximation from \citet{lipari1993} that we verified to be valid for $E_{\text{kin},\mu} > 10$ MeV.

The flux of cosmic rays in the solar system is considered to be isotropic and represented by the Global Spline Fit (GSF) \citep{2017ICRC...35..533D}, which is an modern parameterization of cosmic ray fluxes at Earth between rigidity of a few GV and the highest observed energies at Earth. The rigidity cutoff due to the planetary magnetic field, the location within the solar system and the Sun's level of activity affects the low energy cosmic rays below a few GV/nucleon. The omission of magnetic fields and solar modulation is considered as a source of uncertainty of this calculation but is not expected to qualitatively change the result, in particular not at larger depths or underground.

The absorbed dose rate $D$ in Gy/s is calculated from the differential fluxes $\Phi$ with the default units (GeV cm$^2$ s sr)$^{-1}$ using
\begin{equation}
\label{eq:dose_equation}
\nonumber
    D(X) = 2\pi \sum_{p}\int_{\cos{\theta_{max}}}^1 {\rm d}\cos{\theta}\int {\rm d} E_p~\Phi_p(E_p,X) \left \langle \frac{{\rm d}E_p}{{\rm d}X} \right \rangle (E_p),
\end{equation}
where $\theta_{max} = \max(\pi/2,~\pi - \arcsin{(R/(R + h))}$, $h$ the altitude above ground and $R$ the radius of the planet. The index $p$ iterates over particle species. The expression for $\cos{\theta_{max}}$ takes into account that at higher altitudes, particle cascades can develop upwards relative to the horizon.

The models of the planetary atmospheres are the \citet{us_std_atmosphere} for Earth, derived from Huygens Atmospheric Structure Instrument measurements for Titan \citep{fulchignoni2005}, from the Soviet VeGa-2 probe \citep{1986SvAL...12...40L,lebonnois2017deep} for Venus,  and from a 
fit of piece-wise barometric formulae to Mars Global Surveyor data %\noteAF{Anatoli@Roger: we didn't any suitable parameterization for Mars' atmosphere except \url{https://www.grc.nasa.gov/www/k-12/airplane/atmosmrm.html} The requirement is that the data/parameterization reaches ground and returns density as a function of altitude. There are several papers on the upper atmosphere $> 20$ km but few to none on the troposphere (where most density sits)}. 
Underground fluxes are computed for a water target with the ionization loss computed computed with ESTAR and PSTAR \citep{ionization_tables}. Within the precision of this calculation and depths $<1$ km, the difference in stopping power compared to a dedicated calculation in standard rock is negligible for muons.

\begin{figure}
\centering
\includegraphics[width=.6\textwidth]{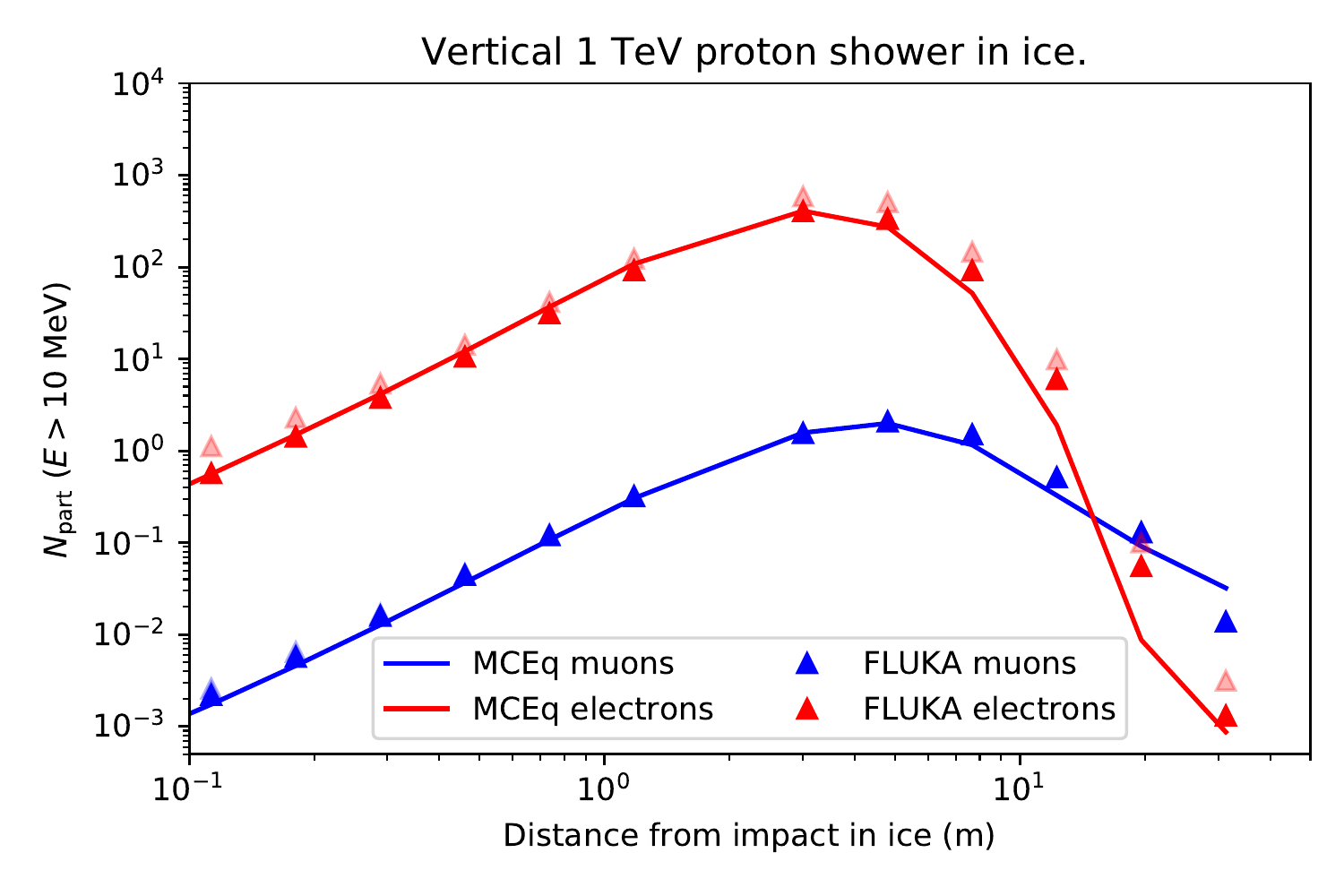}
\caption{Particle number in a particle cascade initiated by a 1 TeV proton in ice. In MCEq the hadronic interaction model is {\sc DPMJET-III-19.1} and {\sc PEANUT} is used as default model in FLUKA 4.0. The full colored triangles represent results for a lower integration cutoff at $E_{\text{kin}, \text{e}^\pm|\mu^\pm} > 10$ MeV whereas the fainter triangles are obtained for cutoff value of  $10$ keV. The difference between the faint and full markers can be used to estimate the fraction particles lost due to the lower grid boundary of 10 MeV in \mceq{}.}
\label{fig:MCEq_vs_FLUKA_NPart}
\end{figure}
\begin{figure}
\centering
\includegraphics[width=.95\textwidth]{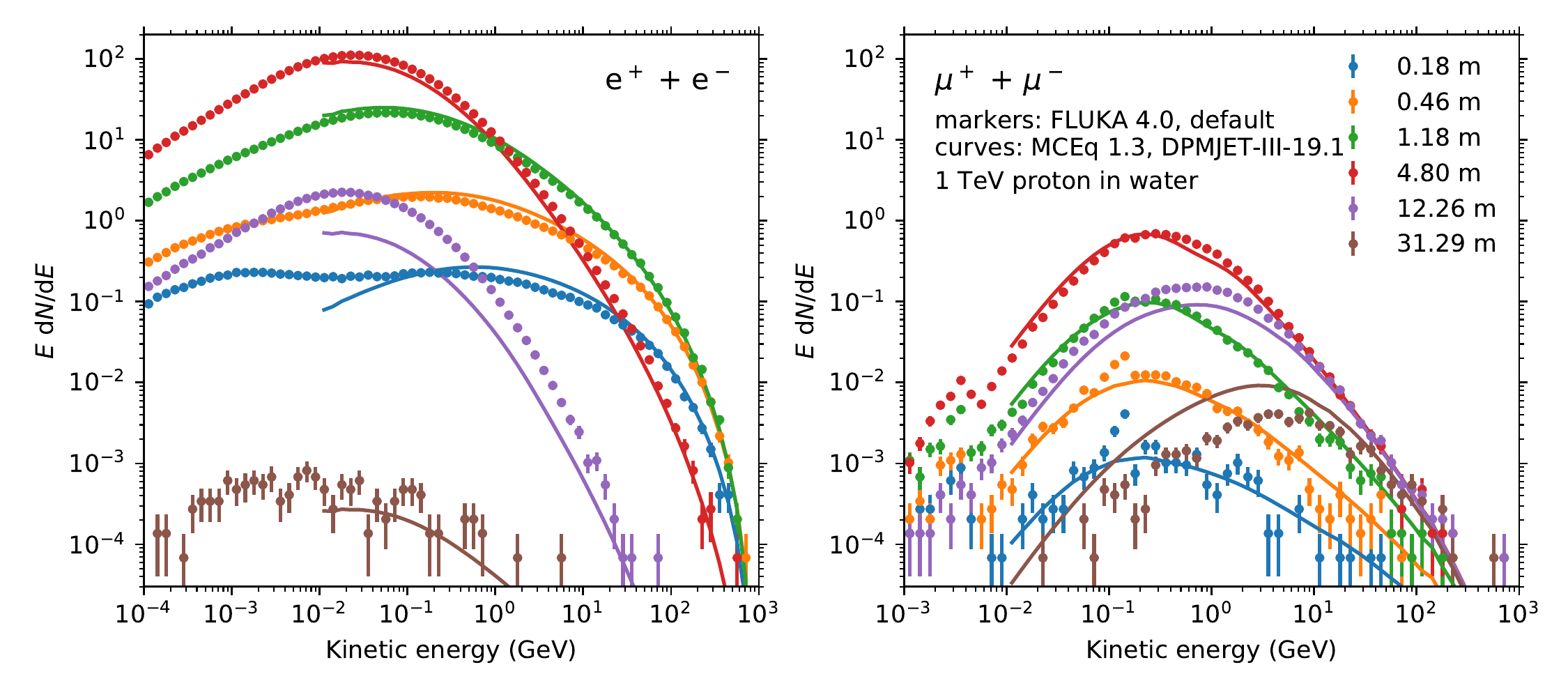}

\caption{Comparison of electron + positron (left) and muon ($\mu^+ + \mu^-$) (right) spectra between \mceq{} and {\sc FLUKA} using the same settings as in Fig.~\ref{fig:MCEq_vs_FLUKA_NPart}. For muon fluxes, the origin of the differences is mostly related to the hadronic interaction model {\sc DPMJET-III-19.1} in \mceq{} vs {\sc PEANUT} in FLUKA. The hadronic model also affects the electromagnetic cascades, but differences at energies below 100 MeV may arise from incomplete   partially the case  For electromagnetic cascades the differences are partly related to,  For very low energy muons, MCEq tends to predict lower fluxes than FLUKA. This is likely related to using {\sc DPMJET-III} at energies close to particle production threshold, where it is known to be incomplete.}
\label{fig:MCEq_vs_FLUKA_spec}
\end{figure}

In Figs.~\ref{fig:MCEq_vs_FLUKA_NPart} and \ref{fig:MCEq_vs_FLUKA_spec} \mceq{} is compared with {\sc FLUKA 4.0}\footnote{\url{https://fluka.cern}} \citep{Battistoni:2015epi} simulations in water for 1 TeV proton projectiles. The technical low energy cutoff in \mceq{} is 10 MeV, which is sufficient to capture a good description for the total number of muons (blue curve in Fig.~\ref{fig:MCEq_vs_FLUKA_NPart} and right panel of Fig.~\ref{fig:MCEq_vs_FLUKA_spec}). At tens of meters in depth, the differences between codes become more pronounced since the particle cascade is increasingly affected by multiple subsequent interactions that amplify differences in production and absorption cross sections between the models. To simulate water targets with \mceq{} it was necessary to include ionization losses for all charged particles and obtain inelastic cross sections for water from the hadronic interaction models. None the less the agreement is remarkably good since \mceq{} has not yet been employed for simulations down to low energies and water targets. For electrons, simulated using \mceq{} and electromagnetic cross sections from \citet{2019ICRC...36..961M}, the agreement is good except at around the cascade maximum. Since the majority of electromagnetic energy is of hadronic origin through the $\pi^0 \to \gamma \gamma$ process, the hadronic model differences between {\sc DPMJET-III-19.1} and {\sc PEANUT/FLUKA} have impact. An estimate for the fraction of electrons that are lost due to the 10 MeV cutoff in \mceq{} can be made from the comparison between the pale and dark red triangles in Fig.~\ref{fig:MCEq_vs_FLUKA_NPart}. Since the gradient of the electromagnetic component is very steep after the maximum, the differences between \mceq{} and {\sc FLUKA} do not qualitatively alter our conclusions from the main text with respect to the depth at which polarized muons constitute the dominant source of cosmic radiation.

\end{document}